\documentclass[epj]{svjour}
\usepackage{amsmath,amsfonts,bm,graphicx,amssymb}
\usepackage{graphics}
\graphicspath{{Figures/}}

\newcommand{\Ham}{\mathcal{H}}
\newcommand{\spin}[1]{\mathbf{S}_{#1}}
\newcommand{\Up}{\uparrow}
\newcommand{\down}{\downarrow}

\newcommand{\elem}[3]{\langle{#1}\vert{#2}\vert{#3}\rangle}
\newcommand{\ket}[1]{\vert{#1}\rangle}
\newcommand{\bra}[1]{\langle{#1}\vert}
\newcommand{\braket}[2]{\langle{#1}\vert{#2}\rangle}

\DeclareMathOperator \Ai {Ai}
\DeclareMathOperator \Bi {Bi}

\DeclareMathOperator \He {\Theta}

\newcommand{\mymat}[1]{\big[{#1}\big]}

\newcommand{\MGstate}{\includegraphics[width=3cm,clip]{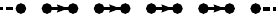}}
\newcommand{\dimer}{\includegraphics[width=4mm,clip]{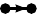}}

\newcommand{\spinon}{\includegraphics[width=2cm,clip]{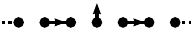}}
\newcommand{\spinonHopeRight}{\includegraphics[width=2cm,clip]{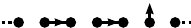}}

\newcommand{\spinonExcited}{\raisebox{-1mm}{\includegraphics[width=2cm,clip]{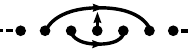}}}

\newcommand{\twospinons}{\includegraphics[width=2.2cm,clip]{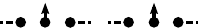}}
\newcommand{\spinonroof}{\includegraphics[width=3.2cm,clip]{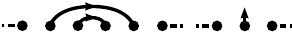}}

\begin{document}

\title{Spinon excitation spectra of the $J_1$-$J_2$ chain from analytical calculations in the dimer basis and exact diagonalization}
\author{Arthur Lavar{\'e}lo \and Guillaume Roux}
\institute{LPTMS, Univ. Paris-Sud and CNRS, UMR 8626, F-91405 Orsay, France.}
\authorrunning{A. Lavar{\'e}lo \and G. Roux}
\titlerunning{Spinon excitation spectra of the $J_1$-$J_2$ chain from analytical calculations in the RVB basis}

\date{\today}
%
\abstract{The excitation spectrum of the frustrated spin-$1/2$
  Heisenberg chain is reexamined using variational and exact
  diagonalization calculations. We show that the overlap matrix of the
  short-range resonating valence bond states basis can be inverted
  which yields tractable equations for single and two spinons
  excitations. Older results are recovered and new ones, such as the
  bond-state dispersion relation and its size with momentum at the
  Majumdar-Ghosh point are found. In particular, this approach yields
  a gap opening at $J_2=0.25J_1$ and an onset of incommensurability in
  the dispersion relation at $J_2=9/17J_1$ [as in S. Brehmer \emph{et
  al.}, J. Phys.: Condens. Matter \textbf{10}, 1103 (1998)]. These
  analytical results provide a good support for the understanding of
  exact diagonalization spectra, assuming an independent spinons
  picture.
} 

\maketitle

Frustration in antiferromagnetic magnets is one of the key ingredient
to stabilize exotic phases~\cite{Lecheminant2004}. In one-dimension,
where quantum fluctuations destroy the N{\'e}el order, a next-nearest
neighbor coupling is known to bring two main features. There is first
a transition from the quasi-long range ordered phase to a gapped phase
which order parameter is the dimerization, breaking translational
invariance. The second is the onset of incommensurability in the spin
correlations and dispersion relation of elementary excitations. The
$J_1$-$J_2$ frustrated chain model is thus a paradigmatic model for
quantum magnetism which has been widely studied and from which stemmed
the physics of valence bond solid phases.

We start by recalling known results on the frustrated spin-$1/2$ chain
Hamiltonian which reads
\begin{equation}
\label{eq:Ham_J1J2}
\Ham=\sum_{i=1}^L J_1\spin{i}\cdot\spin{i+1} +J_2\spin{i}\cdot\spin{i+2}\;,
\end{equation}
in which $J_{1,2}>0$ are antiferromagnetic couplings and $\spin{i}$
are spin-$1/2$ operators. $L$ denotes the length of the chain and
periodic or open boundary conditions can be used.

The phase transition to a dimerized state can be understood by
bosonization arguments, leading to a Kosterlitz-Thouless type of
transition~\cite{Haldane1982,Tonegawa1987,Emery1988}. The transition
point can be efficiently determined by level
spectroscopy~\cite{Okamoto1992} and is found to be located at $J_2/J_1
\simeq 0.241167$~\cite{Tonegawa1987,Okamoto1992,Eggert1996}. One
peculiarity of the transition is the disappearance of logarithmic
corrections associated with SU(2) symmetry right at the critical
point~\cite{Eggert1996}. Another way to understand the opening of the
gap and the onset of a dimerized phase is to start the bosonization
from the limit of two chains coupled in a zig-zag
geometry~\cite{White1996}, ie. the $J_2\gg J_1$ limit. Then, the gap
is shown to decay exponentially with $J_2/J_1$, so as the
dimerization. Between both regimes, the gap and dimerization curves
display a intermediate maximum (not at the same location for both
quantity) which is captured by
numerics~\cite{Tonegawa1987,White1996}. Deep in the dimerized phase,
the two degenerate ground-states in the thermodynamical limit are well
pictured by the Majumdar-Ghosh (MG) state which is a product of
decoupled dimers on bonds $\ket{\text{MG}}=\ket{\MGstate}$, where
dimers are represented by $\ket{\dimer} = \frac{1}{\sqrt{2}}
(\ket{\Up\down} - \ket{\down\Up})$ and an even length is
assumed. Actually, for the special value $J_2=J_1/2$, this MG state
is the exact ground-state of the Hamiltonian
\cite{Majumdar1969,Majumdar1969a}. Similar and generalization of such
valence bond solid states have been found is other models and remain
an active field of research since having exact and simple
ground-states to non-trivial Hamiltonians is essential to the
understanding of quantum magnetism.

The other feature introduced by frustration is incommensurability in
real-space spin correlations: they start to oscillate at a wave-vector
$q\neq \pi$ above a point called the disorder point. This effect
already emerges in the classical limit~\cite{Villain1959} in which it
is simply understood as a best compromise between the two couplings,
one aiming to order at $\pi$ while the other aims to order at
$\pi/2$. The classical disorder point is at $J_2/J_1 = 0.25$. In the
quantum version, the easiest, and actually the only quantitative way
so far, is to compute numerically the real-space spin
correlations~\cite{Tonegawa1987,White1996,Aligia2000,Kumar2010a}. It
can be argued~\cite{Nomura2003} that the onset of incommensurability
should match the minimum of the spin correlation length which
corresponds to the MG point in the model under study. This scenario is
confirmed in the numerics and also supported by
variational~\cite{Zeng1995} and perturbative~\cite{Arovas1992}
arguments.  Moreover, incommensurability will naturally show up in the
spin structure factor then~\cite{White1996,Zeng1995,Bursill1995}. Yet,
in the presence of a finite correlation length, the onset of
incommensurability in this signal occurs for a stronger
frustration~\cite{Nomura2003}, called the Lifshitz point. This point
has been estimated numerically at
$J_2/J_1=0.52036$~\cite{Bursill1995}, and it has been recently shown
that for odd chains (in which the MG state cannot be the
ground-state), the Lifshitz point is actually shifted toward a larger
value $J_2/J_1 = 0.538$.

Last, the dispersion relation becomes incommensurate too, in the sense
that the minimum of the triplet excitation lies at a wave-vector away
from the $K=\pi$ antiferromagnetic wave-vector, and for which we will
use the notation $q^*$ in the following. The connection between
dispersion relation and spin correlations is not physically
straightforward. Still, approximations such as the single-mode
approximation help capture the minimum of the dispersion relation from
correlators~\cite{Arovas1992,Arovas1988}. This approach works best for
gapped systems and was successfully applied to other models known to
display incommensurability, such as the spin one bilinear-biquadratic
chain~\cite{Schollwock1996,Golinelli1999}. In the situation of two
strongly coupled $J_1$-$J_2$ chains, one can even analytically show
that $q^*\neq q$~\cite{Lavarelo2011}. Together with the dispersion
relation, the dynamics of elementary excitations has been investigated
in several limits of the model. For the Heisenberg chain, the
Bethe-ansatz solution provides a quantitative and physically
transparent picture in terms of two-spinons continuum, corresponding
to the so-called des-Cloizeaux Pearson
law~\cite{Cloizeaux1962,Yamada1969,Hashimoto1982}. Multi-spinons or
multi-magnons contributions have recently been shown to be relevant in
the spectral weight of several dynamical structure
factors~\cite{Barnes2003,Caux2006,Mourigal2013}. At the MG point,
variational methods have been used to tackle the dispersion
relation~\cite{Shastry1981,Caspers1984,Frahm1997,Byrnes1999,Uhrig1999,Mkhitaryan2006,Kokalj2010,Deschner2011,Deschner2013}.
Working in the dimer basis gave a good account for the shape of the
dispersion relation, with in particular the explanation for a triplet
bound-state~\cite{Shastry1981} close to $K=\pi/2$. Going away from the
MG point is more difficult and other techniques such as matrix-product
states~\cite{Brehmer1998} have been used to clarify the behavior and
the onset of incommensurability that was found to be at $J_2/J_1 =
9/17 \simeq 0.52941$. This shows that this defines a third different
point for the onset of incommensurability.

In addition to the spontaneously dimerized phase, an explicit
alternating neareast neighbor coupling $J_1+(-1)^i\delta$ will
explicitly break translational invariance and bring the system into
the spin-Peierls phase~\cite{Chitra1995}. The MG point condition can
be generalized in this case and a non-zero $\delta$-term induces the
confinement of spinon
excitations~\cite{Affleck1997,Sorensen1998,Bouzerar1998,Augier1999}. Many
works have been devoted to this physics, using numerical and
analytical
methods~\cite{Byrnes1999,Uhrig1999,Chitra1995,Sorensen1998,Shevchenko1999,Augier1999,Barnes1999,Zheng2001,Hamer2003,Schmidt2004,Takayoshi2012}
to characterize the confinement between two spinons, leading to a
bound-state, as well as the confinement to a chain end for odd length
systems.

In this paper, we study the elementary spinon excitations of the
$J_1$-$J_2$ chain using a variational description well suited for the
MG point and extending the results around it to capture features such
as the dimerization transition and the onset of
incommensurability. The variational method in the resonating valence
bond (RVB) basis has already been used
previously~\cite{Shastry1981,Caspers1984,Byrnes1999,Uhrig1999,Deschner2011,Deschner2013}
and we show that more analytical results can be obtained by a
systematic projection of the Hamiltonian on this basis, rather than
solving numerically the generalized eigenvalue problem, as usually
done. This strategy was first used successfully in the case of a
random MG chain~\cite{Lavarelo2013}. To complete this approach on
elementary excitations, the results are compared to exact
diagonalization (ED) spectra showing a good description of both single
and many spinons excitations.

The paper is organized as follows: we first present the RVB
variational approach and the way one can derive an effective
Hamiltonian from it. The method is then applied to the single spinon
dispersion relation at, and away from the MG point. We then move to
the two-spinon spectrum which is obtained only for the MG point but
with an explicit determination of the triplet bound-state
size. Comparison of ED with an independent spinons ansatz is
given. Last, we discuss the situation of explicit dimerization,
deriving properly the confinement of a single spinon and showing how
to obtain the two-spinon excitation spectrum analytically at the MG
point.

\section{The RVB basis variational method}
\label{sec:variational_approach}

\subsection{Presentation of the method in the case of a single spinon
  excitation}

In order to study the dynamics of a single spinon, one can work on an
odd size chain of length $L$.  We assume open boundary conditions, so
that the spinon lives on one of the two sub-lattice only. Using
periodic boundary conditions would make the spinon jump on the other
sub-lattice while arriving at the end of the chain, corresponding to a
doubling of the number of available sites (roughly doubling the system
size). The variational approach consists in working with the subspace
generated by states of the form $\ket{2i}=\ket{\spinon}$ with a spinon
at site $2i$ ($i\in[0,\frac{L-1}{2}]$) which separates two MG
domains. These states are clearly intuitive for the MG point. They
constitute a free family but they do not span~\cite{Saito1990} the
whole spin sector $\{S_\text{tot}=1/2,\ S^z_\text{tot}=1/2\}$, making
the approach variational. Notice that including the spinon states
living on the other sublattice, as one would do with periodic boundary
conditions, makes the family over-complete and we thus prefer to work
with open boundary conditions. A crucial point is that the states are
non-orthogonal. The overlap matrix $\mymat{\mathcal{O}}$ (in the
following, we use the notation $\mymat{A}$ for the matrix
representation of operator $A$ since Dirac notation can be confusing
while working with non-orthogonal states) has elements
\begin{equation}
\mymat{\mathcal{O}}_{ij} = \braket{2i}{2j}=\left(-\frac{1}{2}\right)^{|i-j|}\;.
\end{equation}

\subsection{Effective Hamiltonian}

The goal of the variational approach is to diagonalize the restriction
$\widetilde{\Ham}$ of $\Ham$ in the subspace $\{\ket{2i}\}$ for which
we have
\begin{equation}
\widetilde{\Ham}=P\Ham P\;,
\end{equation}
where $P$ is the orthogonal projector on that subspace.  Since $P$ is
self-adjoint, the effective Hamiltonian $\widetilde{\Ham}$ is
self-adjoint two which ensures that its eigenvalues are real.
Diagonalizing the representation of $\widetilde{\Ham}$ in the basis
$\{\ket{2i}\}$ is a generalized eigenvalue problem that reads
\begin{equation}
\sum_i\elem{2j}{\Ham}{2i}\psi_i = E\sum_i\braket{2j}{2i}\psi_i\;.
\label{eq:generalized_EVP}
\end{equation}
where $E$ is an eigenenergy and 
\begin{equation}
\label{eq:psi_varia}
 \ket{\psi} = \sum_i \psi_i \ket{2i}
\end{equation}
is the decomposition of the associated eigenfunction in this basis.

We would like to stress the fact that the
$\elem{2j}{\widetilde{\Ham}}{2i}$ are not the matrix elements
$\mymat{\widetilde{\Ham}}_{ij}$ of $\widetilde{\Ham}$ in the basis
$\{\ket{2i}\}$ since the latter in non-orthogonal. Yet, they are
connected by the projector $P$ which writes as the inverse of the
overlap matrix
\begin{equation}
P = \sum_{ij} \mymat{\mathcal{O}^{-1}}_{ij} \ket{2i}\bra{2j}\;.
\end{equation}
Then, one gets the relation
\begin{equation}
\mymat{\widetilde{\Ham}}_{ij} = \sum_k \mymat{\mathcal{O}^{-1}}_{ik}\elem{2k}{\Ham}{2j}\;.
\end{equation}
In general, inverting the overlap matrix is hard and authors prefer to
solve \eqref{eq:generalized_EVP} directly with numerical methods.
However, in the case of a chain, we found that the inverse of the
overlap matrix takes the following tridiagonal form
\begin{equation}
\mymat{\mathcal{O}^{-1}} = \frac 1 3
\begin{pmatrix}
 4 & 2 &   &   &   & \vspace{0.15cm}   \\
 2 & 5 & 2 &   &   & \vspace{0.15cm}   \\
\hspace{0.4cm}   & 2 & 5 & 2 &   &    \\
   &   & \ddots & \ddots & \ddots &    \\
   &   & \vspace{0.15cm} &   2    &   5   & 2   \\
   &   &        &       &    2 &  4 \\
\end{pmatrix}\;,
\end{equation}
which will allow some analytical solutions of the diagonalization of
the effective Hamiltonian.

Notice that we do not specify any particular Hamiltonian so far and
the approach could be used to spin-$1/2$ chain models other than the
frustrated chain. In what follows, we work with \eqref{eq:Ham_J1J2}
which can rewritten at the MG point as
\begin{equation}
\Ham_\text{MG}=J\sum_i\left(2\spin{i}\cdot\spin{i+1}+\spin{i}\cdot\spin{i+2}\right)\;,
\label{eq:H_MG}
\end{equation}
with $J=J_2$. We also recall two basic facts on the two MG
states. Their energy is
\begin{equation}
E_\text{MG}=-\frac34LJ\;,
\label{eq:E_MG}
\end{equation}
for periodic boundary conditions and for the ground-state with open
boundary conditions. Second, they are non-orthogonal and their overlap
reads
\begin{equation}
\braket{\text{MG}}{\text{MG'}}=-\left(-\frac12\right)^{\frac L2-1}\;,
\label{eq:recouvrement_MG}
\end{equation}
irrespective of the boundary conditions.

\section{Dynamics of a single spinon}

\subsection{Dispersion relation at the MG point}
\label{sec:dispersion_spinon}

We first show how the method allows one to recover the spinon
dispersion relation at the MG point. To do so, we look at the
application of the terms in \eqref{eq:H_MG} on a basis state
$\ket{2j}=\ket{\spinon}$ :
\begin{align}
\spin{2j}\cdot\spin{2j+1}\ket{2j} &= +\frac{1}{4}\ket{2j} + \frac{1}{2}\ket{\spinonHopeRight}\;, \label{eq:state1}\\
\spin{2j}\cdot\spin{2j+2}\ket{2j} &= -\frac{1}{4}\ket{2j} - \frac{1}{2}\ket{\spinonHopeRight}\;, \label{eq:state2}\\
\spin{2j-1}\cdot\spin{2j+1}\ket{2j} &= +\frac{1}{4}\ket{2j} + \frac{1}{2}\ket{\spinonExcited}\;. \label{eq:state3}
\end{align}
One can check that the last state \eqref{eq:state3} is orthogonal to the
variational subspace. Then, we deduce the restriction of
$\Ham_{\text{MG}}$ to this subspace
\begin{equation}
\big(\widetilde{\Ham}_{\text{MG}}-E_{\text{MG}}\big)\ket{2j} = \frac{J}{2}\Big(\ket{2j-2}+\frac{5}{2}\ket{2j}+\ket{2j+2}\Big)\;,
\label{eq:Heff_MG}
\end{equation}
where $E_{\text{MG}}$ is given by \eqref{eq:E_MG} extrapolated to odd
sizes.  Thus, we are left with a simple tight-binding Hamiltonian
which is straightforwardly diagonalized by Fourier transformation of
the states
\begin{equation}
\ket{k}=\sum_je^{ik2j}\ket{2j}\;, 
\end{equation}
with $k\in[\frac\pi2,\frac\pi2]$ due to the folding of the Brillouin
zone. Then, the well-known~\cite{Shastry1981} dispersion relation of a
single spinon at the MG point is recovered
\begin{equation}
\omega(k)=J\left(\frac 5 4+\cos 2k\right)\;.
\end{equation}

\subsection{Away from the MG point: single spinon gap and
  incommensurability}
\label{sec:incommensurabilite_J1J2}

We now show that the same approach can be extended away from the MG
point, capturing the main two features of the phase diagram: the
transition to a gapless phase at small $J_2$ and the onset of
incommensurability in the dispersion relation.  This is done by
rewriting \eqref{eq:Ham_J1J2} as
\begin{equation}
\Ham=\Ham_\text{MG}+\eta\sum_i \spin{i}\cdot\spin{i+1}\;,
\end{equation}
with the parametrization $\eta=J_1-2J$ which measures the distance
from the MG point.  The $\eta$-term is simply a nearest-neighbor term
which, when applied to state $\ket{2j}$, generates diagonal terms and,
more importantly, terms with dimer excitations of the form
\begin{align*}
\spin{2i-1}\cdot\spin{2i}\ket{2j}=\frac14\ket{2j}+\frac12\ket{[2i-2,2i+1],2j}\quad(i<j)\;,\\
\spin{2i}\cdot\spin{2i+1}\ket{2j}=\frac14\ket{2j}+\frac12\ket{2j,[2i-1,2i+2]}\quad(i>j)\;,
\end{align*}
in which $ \ket{[2i-2,2i+1],2j}=\ket{\spinonroof}$ is the state having
a spinon at site $2j$ and a singlet between sites $2i-2$ and $2i+1$.
The overlaps of these states with $\ket{2j}$ match
\begin{align*}
\braket{2j}{[2n-2,2n+1],2i}  =-\frac{1}{2}\braket{2j}{2i}(1+3\He(n-j-1))\;,\\
\braket{2j}{2i,[2n-1,2n+2]} =-\frac{1}{2}\braket{2j}{2i}(1+3\He(j-n-1))\;,
\end{align*}
with $\He$ the Heaviside function such that $\He(0)=1$. One then
obtains the projection onto the variational subspace as
\begin{align*}
P\ket{[2n-2,2n+1],2i} = &-\frac{1}{2}\ket{2i}-\Big(-\frac{1}{2}\Big)^{i-n+1}\ket{2n} \\
 &+\Big(-\frac{1}{2}\Big)^{i-n}\ket{2n-2}\;, \nonumber\\
P\ket{2i,[2n-1,2n+2]} = &-\frac{1}{2}\ket{2i}-\Big(-\frac{1}{2}\Big)^{n-i+1}\ket{2n} \\
 &+\Big(-\frac{1}{2}\Big)^{n-i}\ket{2n+2}\;. \nonumber
\end{align*}
Thus, the correction to the MG Hamiltonian creates arbitrary long
hoppings of dimers which amplitude decreases exponentially with
distance. Finally, the effective Hamiltonian is a full matrix for
which we have
\begin{align}
\big(\widetilde{\Ham}-E_0\big)\ket{2j}&=\Big(\frac54J+\frac78\eta\Big)\ket{2j} \label{eq:Heff_J1J2}\\
&+\frac12(J+\eta)\Big(\ket{2j-2}+\ket{2j+2}\Big)\nonumber \\
&+\eta\sum_{n<j}\Big(-\frac{1}{2}\Big)^{j-n}\Big(\frac{1}{4}\ket{2n}+\frac{1}{2}\ket{2n-2}\Big)\nonumber \\
&+\eta\sum_{n>j}\Big(-\frac{1}{2}\Big)^{n-j}\Big(\frac{1}{4}\ket{2n}+\frac{1}{2}\ket{2n+2}\Big)\;,\nonumber
\end{align}
where we define $ E_0=E_\text{MG}-\frac{3}{8}\eta L$ as the origin of
energies.  Fourier transforming this relation gives the dispersion
relation
\begin{align}
\omega(k)= & \frac54J+\frac78\eta+(J+\eta)\cos(2k)\\
  &+\frac\eta2\sum_{n>0}\Big(-\frac12\Big)^n\left(\cos\left(2kn\right)+2\cos\left[2k(n+1)\right]\right)\;.\nonumber
\end{align}
The summation can be carried out, leading to the following compact
form
\begin{align}
\omega(k)=&\frac78J_1-\frac12J_2+(J_1-J_2)\cos2k \nonumber\\
      &+4(J_1-2J_2)\frac{\sin^2 2k}{5+4\cos2k}\;.\label{eq:dispersion_J1J2}
\end{align}
The same dispersion relation, though written differently and with a
different constant, was obtained from a matrix-product states ansatz
in Ref.~\cite{Brehmer1998}. Their gap was not the same because the
constant factor is different. Within our variational approach, we
obtain a vanishing gap for $J_2=0.25J_1$ which is actually rather
close to the numerical value for the transition, thereby giving a
simple picture for the onset of the transition, starting from the
dimerized phase. We will come back to this point in
Section~\ref{sec:etats_2spinons}.

The minimum of the dispersion relation is at $k=\pi/2$ for small $J_2$
but becomes incommensurate above the threshold $J_2/J_1 = 9/17 \simeq
0.52941$. This value is the same as in Ref.~\cite{Brehmer1998} and in
agreement with the recent numerical study of
Ref.~\cite{Deschner2013}. In the incommensurate regime, the position
$k^*$ of the minimum of the single-spinon dispersion relation follows
\begin{equation}
k^*=\frac12\arccos\left(-\frac54+\frac34\sqrt{2-\frac{J_1}{J_2}}\right)\;.
\end{equation}
Close to the threshold, the wave-vector thus displays a discontinuity
in its derivative with an exponent $1/2$ in the distance from
commensurability:
\begin{equation}
\label{eq:q*_J1J2}
\frac\pi2-k^*\propto\left(\frac{J_2}{J_1}-\frac{9}{17}\right)^{1/2}\;.
\end{equation}
A similar square-root behavior was obtained in the frustrated ladder
case~\cite{Lavarelo2011} and is compatible with the generic scenario
for the onset of incommensurability~\cite{Nomura2003}.

\begin{figure*}[t]
\hspace{-6mm}
\includegraphics[width=1.03\linewidth,clip]{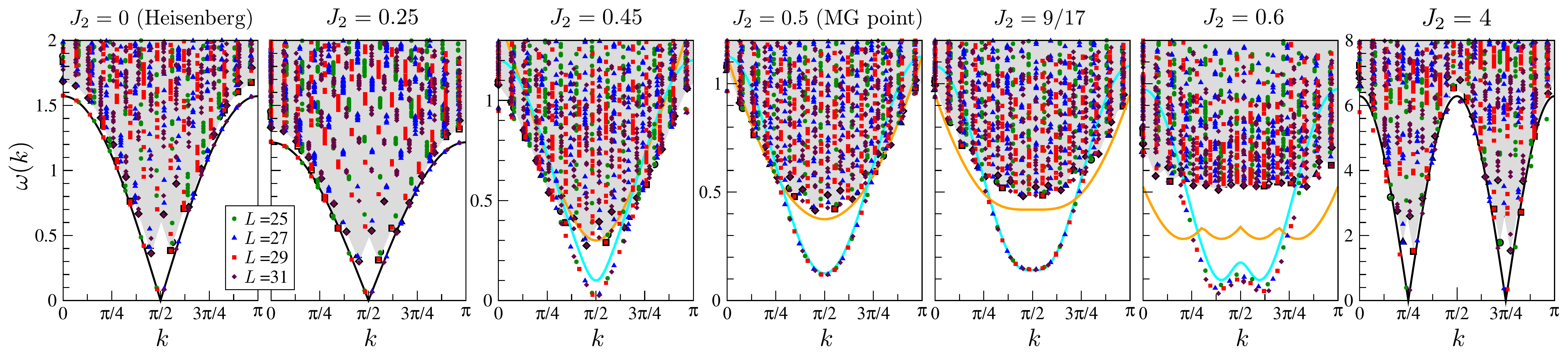}
\caption{\textbf{Single spinon dispersion relations:} variational
  predictions are plotted against exact diagonalization data on odd
  size chains for increasing frustration $J_2$, taking $J_1=1$ as the
  unit of energy. The grey area represents the region for which
  energies belong to the $S_z^{\text{tot}}=3/2$ sector in ED.  The
  black curves are the lowest branches of spinon dispersion expected
  from Bethe ansatz calculations. For $J_2=0$, the $J_1\frac \pi
  2|\cos k|$ is exact while for $J_2=0.25J_1$, the prefactor is fitted
  to the curve. In the $J_2=4J_1$ plot, $J_2\frac \pi 2|\cos 2k|$ is
  used as the system consists in almost decoupled chains with double
  lattice spacing. Notice that, strictly speaking, plots with
  $J_2=0.25J_1$ and $J_2=4J_1$ correspond to gapped systems but the
  gap is so small that the ED data cannot resolve it and the Bethe
  ansatz curve reproduce well the higher energies part.  The cyan
  curve is the variational result \eqref{eq:dispersion_J1J2}.  The
  orange curve is the lowest branch of 3-spinons dispersion relation
  in the variational approach and assuming independent spinons, to be
  compared to the region of states with $S^z_\text{tot}=3/2$.}
\label{fig:incommensurabilite_J1J2}
\end{figure*}

\subsection{Evolution of the dispersion relation with frustration}

With these results, we can sketch the overall behavior of the single
spinon dispersion relation with increasing frustration. Exact
diagonalization spectra have been computed using the Lanczos algorithm
on a symmetrized Hilbert space in both the $S^z_\text{tot}=1/2$ and
$S^z_\text{tot}=3/2$ sectors to observe the opening of the gap and the
onset of incommensurability. The single spinon spectrum at the MG
point was given in Ref.~\cite{Sorensen1998}. The results are displayed
on Fig.~\ref{fig:incommensurabilite_J1J2} for seven typical values of
$J_2/J_1$. First, we recall the physics of the Heisenberg point
($J_2=0$) which is exactly solvable by Bethe ansatz and for which the
lowest spinon branch reads $\omega(k) \simeq J_1\frac {\pi}{2}|{\cos
k}|$ (black curve). For $J_2=0.25J_1$, finite-size effects prevent ED
from a clear determination of the tiny gap that exists in the
thermodynamical limit, and we observe that the higher energies of the
lowest spinon are well reproduced by the Bethe ansatz (or spin-wave)
form with a prefactor smaller than $\pi/2$. In both cases, the
three-spinons energies collapse on the single spinon energy.

Entering deep in the dimerized phase for $J_2=0.45J_1$, we observe the
finite gap in the dispersion relation, and the corresponding quadratic
behavior around the $k=\pi/2$ minimum. Yet, the gap is overestimated
by the variational approach (cyan line) but the shape of the curve is
well reproduced. Using the variational approach and assuming three
independent spinons, one obtains the lowest part of the
$S^z_\text{tot}=3/2$ sector from
\begin{equation}
\omega_{3\text{-spinons}}(k) = \min_{\substack{k_1,k_2,k_3\\\sum_ik_i=k}} \left[\omega(k_1)+\omega(k_2)+\omega(k_3)\right]\;,
\end{equation}
in which $\omega(k)$ is \eqref{eq:dispersion_J1J2} and displayed as
the orange curve on Fig.~\ref{fig:incommensurabilite_J1J2} and which
can be compared to the boundary of the grey area. Naturally, the
three-spinons spectrum is separated from the single spinon branch at
low energies because of the gap. Interestingly, for $k\sim
\pi/4,3\pi/4$, the two dispersion relations touch (even cross) each
other, telling that the single spinon branch enters the continuum of
many spinons. This seems to be in agreement with the ED which shows
that the spinon branch separates from the continuum only for
$k\in[3\pi/8,5\pi/8]$. These main features are recovered at the MG
point, when $J_2=0.5J_1$, with a better agreement between the numerics
and the variational approach, as expected. One reaches the
incommensurate point for $J_2=9/17J_1$ and the ED data display a
flattened dispersion relation signaling the onset of
incommensurability. The gap is large and separates well the spinon
branch from the three-spinon continuum although the branch clearly
enters the continuum on the sides. In the incommensurate regime for
$J_2=0.6J_1$, the dispersion has a double minimum which shape and
position are well reproduce by the variational approach. Yet, the
independent spinons picture fails to account for the large gap between
the branch and the continuum, which may be due to strong spinon
scattering in this regime and the fact that the variational approach
becomes less reliable away from the MG point. Last, in the limit of
large $J_2$, the system is almost equivalent to two decoupled chains,
which corresponds to a doubling of the unit cell. Although there is a
tiny gap for $J_2=4J_1$, the spectrum is again well reproduced by the
Bethe ansatz curve which is $J_2\frac {\pi}{2}|{\cos(2k)}|$ in this
case.

\section{Two spinons excitation spectrum}
\label{sec:etats_2spinons}

\subsection{Variational approach}

We now study the elementary excitations on an even length chain at the
MG point. They correspond to two-spinons excitations and the natural
subspace to describe these excitations is spanned by states of the
form $\ket{x_1,x_2}=\ket{\twospinons}$ ($x_1<x_2$) such that the spin
sector is determined by the two spinons while the rest of the chain is
in the singlet sector because of MG domains. Thanks to spin rotational
invariance, we stick to the $S^z_{\text{tot}}=1$ state for the triplet
sector and otherwise to the singlet state. Again, these states
constitute a free family that does not span the whole singlet or
triplet sectors and they are non-orthogonal. Open boundary conditions
are used and the chain is assumed to be infinite in both
directions. In this situation, spinons cannot change sublattices so
that $x_1$ remains even and $x_2$ odd.

\subsubsection{Schr\"odinger equation}

As long as spinons do not belong to neighboring sites ($x_1+3\leq
x_2$), the effective Hamiltonian will act separately on each spinon,
according to \eqref{eq:Heff_MG}, and independently of the spin sector
\begin{align}
\big(\widetilde{\Ham}_\text{MG}-E_\text{MG}\big)\ket{x_1,x_2}&=\frac52J\ket{x_1,x_2} \\
&+\frac J2\big(\ket{x_1-2,x_2}+\ket{x_1+2,x_2}\nonumber\\
&\qquad+\ket{x_1,x_2-2}+\ket{x_1,x_2+2}\big)\;.\nonumber
\end{align}
One then exploits translational invariance by moving to the center of
mass frame, introducing variables:
\begin{equation}
X=\frac{x_1+x_2}{2}\;\text{ and }\;x=x_2-x_1\;,
\end{equation}
in which $x$ takes odd positives values and $X$ half-integer values.
For a given $x$, all values of $X$ are not allowed. One must have
\begin{equation}
 x=4i\pm1 \Leftrightarrow X=2j\pm\frac12\;.
\end{equation}
Using a Fourier transform on the $X$ coordinate gives states
\begin{equation}
\ket{K,x=4i\pm1}=\sum_{X=2j\pm\frac12} e^{iKX} \ket{X,x}\;,
\end{equation}
in which the first Brillouin zone is $K\in[-\frac\pi2,\frac\pi2]$
since the center of mass jumps by two sites for given $x$. The
effective Hamiltonian then takes the following form
\begin{align}
\big(\widetilde{\Ham}_{\text{MG}}-E_{\text{MG}}\big)\ket{K,x}=&\frac52J\ket{K,x}\\
&+J\cos K\left(\ket{K,x-2}+\ket{K,x+2}\right),\nonumber
\end{align}
when $x\geq3$. The variational wave-function reads
\begin{equation}
\ket{\psi}={\sum_{i\in\mathbb N}}\int_{-\frac\pi2}^{\frac\pi2} \frac{dK}{\pi}\psi_i(K)\ket{K,x=2i+1}\;,
\end{equation}
which yields to the Schr\"odinger equation for $i>1$:
\begin{align}
\left(E-E_\text{MG}\right)\psi_i(K)=&\frac52J\psi_i(K)\label{eq:schrodinger_MG}\\
&+J\cos K\left[\psi_{i-1}(K)+\psi_{i+1}(K)\right]\;.\nonumber
\end{align}

\subsubsection{Singlet sector}

We now have to discuss the case of neighboring spinons which
corresponds to the $x=1$ boundary conditions. There, singlet and
triplet sectors behave differently which will be responsible for a
bound state. 

\begin{figure}[tb]
\centering
\includegraphics[width=0.8\linewidth,clip]{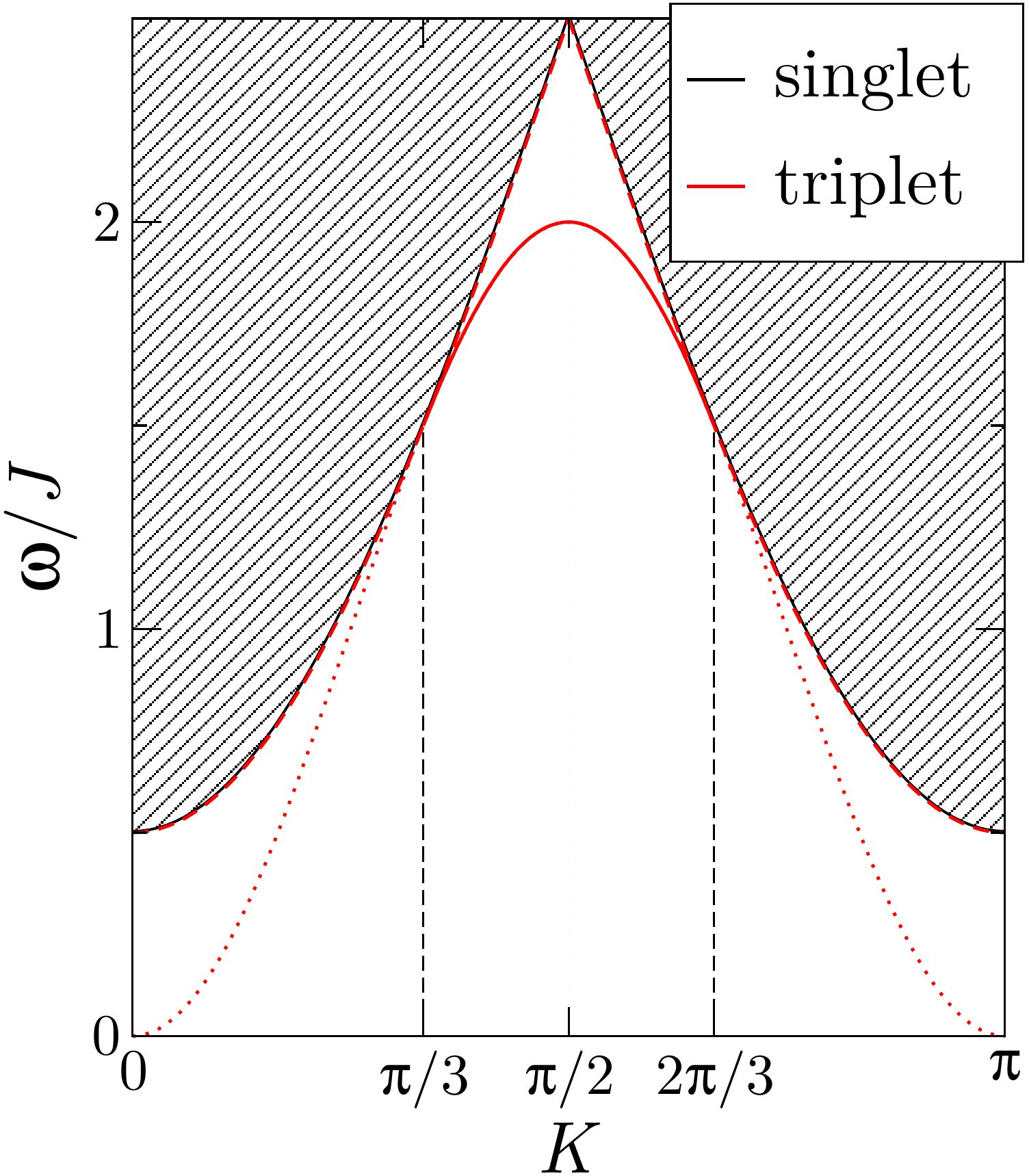}
\caption{Low-energy spectrum of the frustrated chain at the MG
  point. Dashed lines show the continuation of the triplet dispersion
  relation \eqref{eq:dispersion_triplet} for which no state exists.}
\label{fig:dispersion_MG}
\end{figure}

In the singlet sector, the state $\ket{X,x=1}$ actually equals the MG
state for all $X$ and is an eigenstate of energy $E_\text{MG}$. Thus,
in the singlet sector, the boundary terms of the Schr\"odinger
equation read
\begin{align}
\label{eq:limite_sg_1}
&\left(E-E_\text{MG}\right)\psi_1(K)=\frac52J\psi_1(K)+J\cos(K)\psi_{2}(K)\;,\\
\label{eq:limite_sg_2}
&\left(E-E_\text{MG}\right)\psi_0(K)=J\cos(K)\psi_{1}(K)\;.
\end{align}
There are two cases: either $\psi_0(K)\neq 0$ or $\psi_0(K) = 0$.  If
$\psi_0(K)\neq 0$, as $\ket{K,x=1}$ is already an eigenstate with
eigenvalue $E_\text{MG}$, the only possible energy is
$E_\text{MG}$. Then, from \eqref{eq:limite_sg_2} we have
$\psi_{1}(K)=0$ and from \eqref{eq:limite_sg_1} we have
$\psi_{2}(K)=0$. It follows from \eqref{eq:schrodinger_MG} that
$\psi_{i}(K)=0$ for $i \neq 0$ and we thus recover $\ket{\text{MG}}$.
If $\psi_0(K)=0$, in which case \eqref{eq:limite_sg_1} is equivalent
to \eqref{eq:schrodinger_MG} for $i=1$ to which one must add the
boundary condition
\begin{equation}
\psi_0(K)=0\;.
\label{eq:limite_sg}
\end{equation}
Solving \eqref{eq:schrodinger_MG} with the condition
\eqref{eq:limite_sg} yields plane waves of momentum $k$ as the only
possible solutions for the relative motion.  Thus a continuum is
obtained, displayed on Fig.~\ref{fig:dispersion_MG} and which follows
\begin{equation}
\omega(K,k) = \frac52 J + 2J\cos(K)\cos(2k)\;,
\label{eq:dispersion_MG}
\end{equation}
where $k\in[-\frac\pi2,\frac\pi2]$ and $K\in[-\frac\pi2,\frac\pi2]$
(displayed over the range $[0,\pi]$ on Fig.~\ref{fig:dispersion_MG}).

\subsubsection{Triplet sector}

In the triplet sector, the action of the effective
$\widetilde{\Ham}_{\text{MG}}$ on a contact state $\ket{x,x+1}$ is
\begin{align}
\big(\widetilde{\Ham}_{\text{MG}}-E_{\text{MG}}\big)\ket{x,x+1}&= 3J\ket{x,x+1}\\
+&\frac J2\ket{x-2,x-1}+\frac J2\ket{x+2,x+3} \nonumber\\
+&J\ket{x-2,x+1}+J\ket{x,x+3}\;, \nonumber
\end{align}
which, once in the center of mass frame and Fourier transformed, gives
the equation
\begin{equation}
\big(\widetilde{\Ham}_{\text{MG}}-E_{\text{MG}}\big)\ket{K,1}=J(3+\cos 2K)\ket{K,1}+2J\cos K\ket{K,3}\;.
\end{equation}
We then deduce the boundary conditions in the triplet sector
\begin{align}
\label{eq:limite_tp1}
(E-E_\text{MG})\psi_1(K)= & \frac52J\psi_1(K) \\
& \nonumber +J\cos(K) \psi_2(K)+2J\cos(K) \psi_0(K)\;,\\
\label{eq:limite_tp2}
(E-E_\text{MG})\psi_0(K) & =J(3+\cos2K)\psi_0(K)\\
& \nonumber +J\cos(K) \psi_1(K)\;.
\end{align}
One last trick is to change variables
$\psi_0(K)\rightarrow\psi_0(K)/2$ so that equation
\eqref{eq:limite_tp1} gives back Schr\"odinger's equation
\eqref{eq:schrodinger_MG} with the new boundary condition
\begin{equation}
\label{eq:limite_tp}
(E-E_\text{MG})\psi_0(K)=J(3+\cos2K)\psi_0(K)+2J\cos(K) \psi_1(K)
\end{equation}
replacing \eqref{eq:limite_tp2}. The boundary condition
\eqref{eq:limite_tp} still possesses plane wave solutions which gives
back the continuum of diffusion \eqref{eq:dispersion_MG} which is then
four times degenerate.

\begin{figure*}[t]
\hspace{-6mm}
\includegraphics[width=1.03\linewidth,clip]{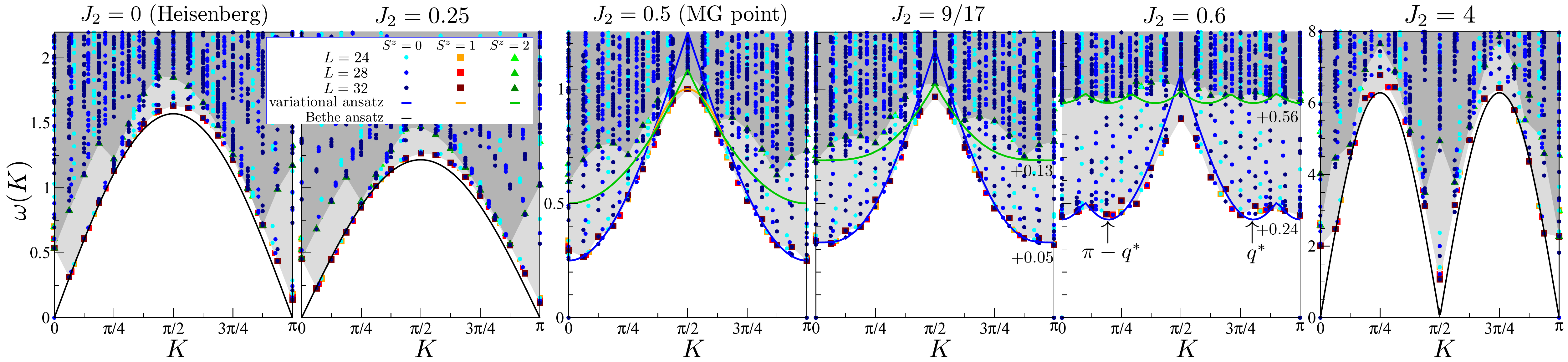}
\caption{\textbf{Two spinons dispersion relations:} Evolution of the
  low-energy spectrum of even size $J_1$-$J_2$ chain for increasing
  $J_2/J_1$, taking $J_1=1$ as the unit of energy. The light grey area
  stands for energy belonging to the $S^z_{\text{tot}}=1$ sector (2
  spinons in a triplet state) while the dark grey area corresponds to
  the $S^z_{\text{tot}}=2$ (4 spinons). These are compared to Bethe
  ansatz (black lines) and the variational approach for two
  independent spinons (blue line, exact at the MG point) and four
  independent spinons (green line). The yellow line at the MG point
  shows the variational result for the triplet bound-state. Extra
  numbers close to the variational lines for $J_2/J_1=9/17,0.6$
  indicate that a vertical shift has been applied to the variational
  prediction to better fit the ED data. }
\label{fig:TwoSpinonSpectra}
\end{figure*}

\begin{figure*}[t]
\hspace{-6mm}
\includegraphics[width=1.03\linewidth,clip]{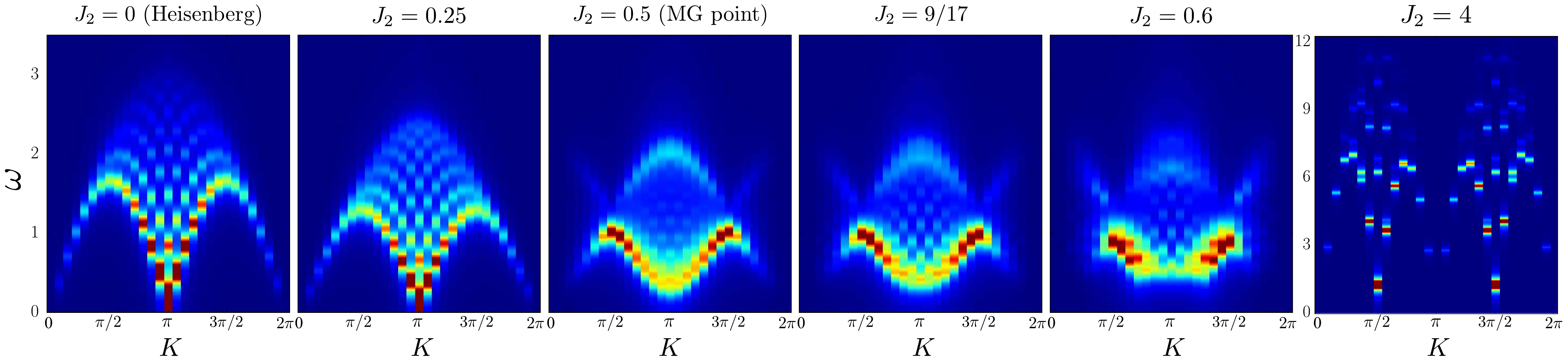}
\caption{\textbf{Dynamical structure factor:} Evolution of the
  dynamical structure factor $\mathcal{S}(K,\omega)$ on an even size
  $J_1$-$J_2$ chain for the same increasing values of $J_2/J_1$ as in
  Fig.~\ref{fig:TwoSpinonSpectra}. Data are obtained from a Lanczos
  calculation on a $L=28$ sites chain.}
\label{fig:SkSpectra}
\end{figure*}

However, the boundary condition \eqref{eq:limite_tp} now allows to
consider bound-state solutions of \eqref{eq:schrodinger_MG}. We then
obtain a triplet state (three times degenerate) with the following
wave-function
\begin{equation}
\psi_i(K)=\phi(K)e^{-2i/\xi(K)}\;,
\end{equation}
in which $\xi(K)$ is the typical width of the spinon bound-state. 
From \eqref{eq:limite_tp}, the dispersion relation reads
\begin{equation}
\label{eq:dispersion_triplet}
\omega(K)=\frac52 J- J\sqrt{4\cos^2K+\left(\frac12+\cos2K\right)^2}\;,
\end{equation}
for $K\in[\frac\pi3,\frac{2\pi}{3}]$, so that the bound-state is below
the singlet continuum, as illustrated on
Fig.~\ref{fig:dispersion_MG}. Notice that for $K=\pi/2$, the
bound-state boils down to $\ket{K=\frac\pi2,x=1}$, which is a
delocalized triplet shared by two neighboring sites. It can be read
directly on \eqref{eq:Heff_MG} that this state is an eigenstate.
Last, this calculation provides the behavior of $\xi_K$ as a function
of the momentum:
\begin{equation}
e^{-2/\xi(K)}=\frac{\sqrt{4\cos^2K+\left(\frac12+\cos2K\right)^2}-\frac12-\cos2K}{2\cos K}\;,
\end{equation}
which displays a divergence $\xi(K) \rightarrow +\infty$ when
$K\rightarrow\frac\pi3^+$ for which one recovers deconfined
excitations as for the singlet sector and the disappearance of the
bound-state.

\subsection{Evolution of the two-spinon spectrum with increasing frustration}

As for the single spinon dispersion relation, we now compare these
predictions to numerics from exact diagonalization on
Fig.~\ref{fig:TwoSpinonSpectra}. In particular, we show the bottom of
the $S^z_{\text{tot}}=1$ and $S^z_{\text{tot}}=2$ spectra to discuss
the bound state and many-spinons excitations. We could not derive the
exact dispersion relation analytically for two spinons away from the
MG point, but the single spinon dynamics can already give insights
through an independent spinons picture in which the two-spinon dispersion
relation reads
\begin{equation}
\omega_{2\text{-spinons}}(K) = \min_{\substack{k_1,k_2\\k_1+k_2=K}} \left[\omega(k_1)+\omega(k_2)\right]\;,
\label{eq:2spinons-ansatz}
\end{equation}
in which $\omega(k)$ is \eqref{eq:dispersion_J1J2}, and for which
singlet and triplet sector are naturally degenerate. Similarly, a four
independent spinons dispersion relation can be obtained to compare
with numerics.

Starting from the Heisenberg point $J_2=0$ the finite-size curve
agrees well with the Bethe ansatz lower branch
$\omega_{2\text{-spinons}}(K)= J_1 \frac{\pi}{2} \sin K$, and for
$J_2=0.25J_1$ again with a slightly smaller prefactor and the fact
that the tiny gap is not captured on finite-size chains. Notice that
in the gapless regime, multi spinons excitations with more than two
spinons are as well gapless in the thermodynamical limit which shows
up the $S^z_{\text{tot}}=2$ sector that gets close to the magnon
branch.

At the MG point, the dispersion relations for the different spin
sectors are very well reproduced by the variational approach. In
particular, the triplet bound-state is the lowest $K=\pi/2$
energy. Interestingly, displaying the $S^z_{\text{tot}}=2$ sector
shows that the four spinons continuums start much higher in energy
than the two-spinon continuum, essentially because of the single
spinon gap, and comes with an increased density of states in the dark
grey region. The independent spinons picture shows that close to and
at $K=\pi/2$, the lowest boundary of the continuum is actually due to
four spinons excitations and not due to two-spinons which lie higher
in energies. It also explains why the triplet bound-state hardly
detaches from the continuum around $K=\pi/2$, while considering only
two-spinon excitations as in Fig.~\ref{fig:dispersion_MG} suggested
that the bound-state would better separate.

Increasing further the frustration ($J_2=9/17J_1$ and $J_2=0.6J_1$)
brings incommensurability in the dispersion relation. The shapes of
the continua are well reproduced by the variational method but, as we
saw for a single spinon, the gap is not quantitatively
reproduced. Therefore, we shifted the variational curves upward by an
amount which is specified on the two plots for each line. The shift
for the four spinons line is nearly twice the one for the two spinons
line which is reasonable. For $J_2=9/17J_1$, the flattening to a
quartic behavior for small $K$ is visible in the ED data thanks to the
variational curve. The higher part of the spectrum is very similar to
the MG point, with the triplet bound-state that still detaches from
the continuum by a minute amount. For $J_2=0.6J_1$, the incommensurate
wave-vector $q^*$ is well resolved by the ED data and in good
agreement with the variational prediction. Looking for the minimum
over $K$ in \eqref{eq:2spinons-ansatz} gives twice the minimum of the
single spinon dispersion relation. Thus, the independent spinon
picture gives the following prediction for the incommensurate
wave-vector:
\begin{equation}
q^*=\arccos\left(-\frac54+\frac34\sqrt{2-\frac{J_1}{J_2}}\right)\;.
\end{equation}
Interactions between spinons can affect both the magnitude of the gap
and the incommensurate wave-vector through their variation with the
relative distance. At the MG point, this dependence is very weak at
small $k$ since the interaction is essentially local. As the
incommensurability in $q^*$ occurs close to the MG point, these
corrections should be small. The least one can say is that ED cannot
resolve them. Notice that both the ED and the variational ansatz shows
that the $K=0$ energy is actually almost degenerate with the $K=q^*$
energy. This may be seen as a precursor of the degeneracy of the
ground-state in the large $J_2$ limit in which $q^*\rightarrow \pi/2$
gets degenerate with the $K=0$ sector. Interestingly and as expected
from an independent spinon picture, the bottom of the four spinons
continuum also displays incommensurability with six minima over the
full Brillouin zone.

Last, in the limit of large $J_2$ and eluding the resolution of the
tiny gap of the system, the excitation spectrum is here again well
fitted by the Bethe ansatz prediction replacing $J_1$ by $J_2$ and
folding the first Brillouin zone (see plot for $J_2=4J_1$).  In
Fig.~\ref{fig:SkSpectra}, the overall process of gap opening and
incommensurate dispersion relation connecting the two limits of a
single chain to two chains is illustrated through the experimentally
accessible spin dynamical structure factor
$\mathcal{S}(K,\omega)$. Lanczos calculations capture a strong
redistribution of the spectral weight close to the MG point, with a
maximum weight around $K\simeq \pi/2$ and a secondary peak at the
higher energies of the two-spinons continuum.

\section{Variational approach for the chain with explicit
  dimerization}
\label{sec:dimerisation_explicite}

In this section, we consider the situation with an explicit
dimerization term $\delta$ for which the Hamiltonian reads
\begin{equation}
\Ham= \sum_i(J_1+(-1)^i\delta)\spin{i}\cdot\spin{i+1} +J_2\spin{i}\cdot\spin{i+2}\;.
\end{equation}
Putting ourselves on the Shastry-Sutherland line $\delta+2J_2=J_1$
\cite{Shastry1981}, the Hamiltonian is rewritten as
\begin{equation}
\Ham=\sum_i(2J+\delta)\spin{i}\cdot\spin{i+1} +J\spin{i}\cdot\spin{i+2}+(-1)^i\delta\spin{i}\cdot\spin{i+1}\;.
\end{equation}
The ground-state of this Hamiltonian is the MG state with dimers on
the strongest bonds $(2i,2i+1)$.  Its energy is
\begin{equation}
 \frac{E_{\text{MG}}}{L}=-\frac34\left(J+\delta\right)\;.
\end{equation}
The other MG is no longer an eigenstate and its energy per site
remains $-\frac34J$ to first order in $\delta$.

\subsection{Spinon confinement on a wall}
\label{sec:confinement}

Let us consider an open chain of odd length starting with a weak link
$J_1-\delta$. Due to the lift of the degeneracy, the explicit
dimerization generates a confinement of the spinon close to the
boundary through a potential linear with the distance $\propto \delta
i$~\cite{Affleck1997,Sorensen1998}. This effect has been studied
analytically in the continuum limit \cite{Byrnes1999,Uhrig1999} and
with DMRG \cite{Sorensen1998}. This section provides a proper
derivation of the continuum limit based on the RVB basis approach. The
explicit dimerization term also generates longer distance hoppings of
the spinon, within the variational picture, inducing
incommensurability above the Shastry-Sutherland line
$\delta>J_1-2J_2$. The latter kinetic effect can be neglected when
$\delta\ll J$ and the confining potential remains the dominant
effect. Within this approximation and choosing a variational
wave-function that mimics the $k=\pi/2$ oscillation corresponding to
the minimum of the dispersion relation \eqref{eq:dispersion_MG}
\begin{equation}
 \ket{\psi}=\sum_i(-1)^i\psi_i\ket{2i+1}\;,
\end{equation}
the Schr\"odinger equation for the spinon reads
\begin{equation}
\left(E-E_{\text{MG}}\right)\psi_i=\left(\frac{5}{4}J+\frac32\delta i\right)\psi_i-\frac{J}{2}\left(\psi_{i-1}+\psi_{i+1}\right)\;,
\end{equation}
to which we add the boundary condition $\psi_{-1}=0$. This equation
can be solved in the continuum limit \cite{Uhrig1999,Byrnes1999},
assuming that $\psi_i$ varies slowly enough with $i$.
In the continuum limit, one has
\begin{equation}
-2J\psi''(x)+\frac{3\delta}4 (x-\varepsilon)\psi(x)=0\;,
\end{equation}
in which $\psi(x=2i+1)=\psi_i$ and $\varepsilon=\frac{4}{3\delta}\left(E-E_\text{MG}\right)-\frac{J}{3\delta}$, that is equivalent to
\begin{equation}
\label{eq:equadiff-Airy}
\psi''(y)-y\psi(y)=0\;,
\end{equation}
after changing variables as follows
\begin{equation}
y=\frac{1}{\xi_\text{conf}}\left(x-\varepsilon\right)\;,\quad\text{with}\quad
\xi_\text{conf}=\left(\frac{8J}{3\delta}\right)^{1/3}\;.
\end{equation}
$\xi_\text{conf}$ is the typical confinement length.  In the limit of
small $\delta\ll J$, one gets $\xi_\text{conf}\gg1$, so that taking
the continuum limit is justified.  A natural set of solutions for the
differential equation \eqref{eq:equadiff-Airy} are Airy functions
$\Ai$ and $\Bi$ that oscillate for $x<0$ and with asymptotic behaviors
\begin{equation}
\Ai(x)\underset{x\rightarrow+\infty}{\longrightarrow}0\;,\quad \Bi(x)\underset{x\rightarrow+\infty}{\longrightarrow}+\infty\;.
\end{equation}
\begin{figure}[t]
\centering
\includegraphics[width=0.9\linewidth,clip]{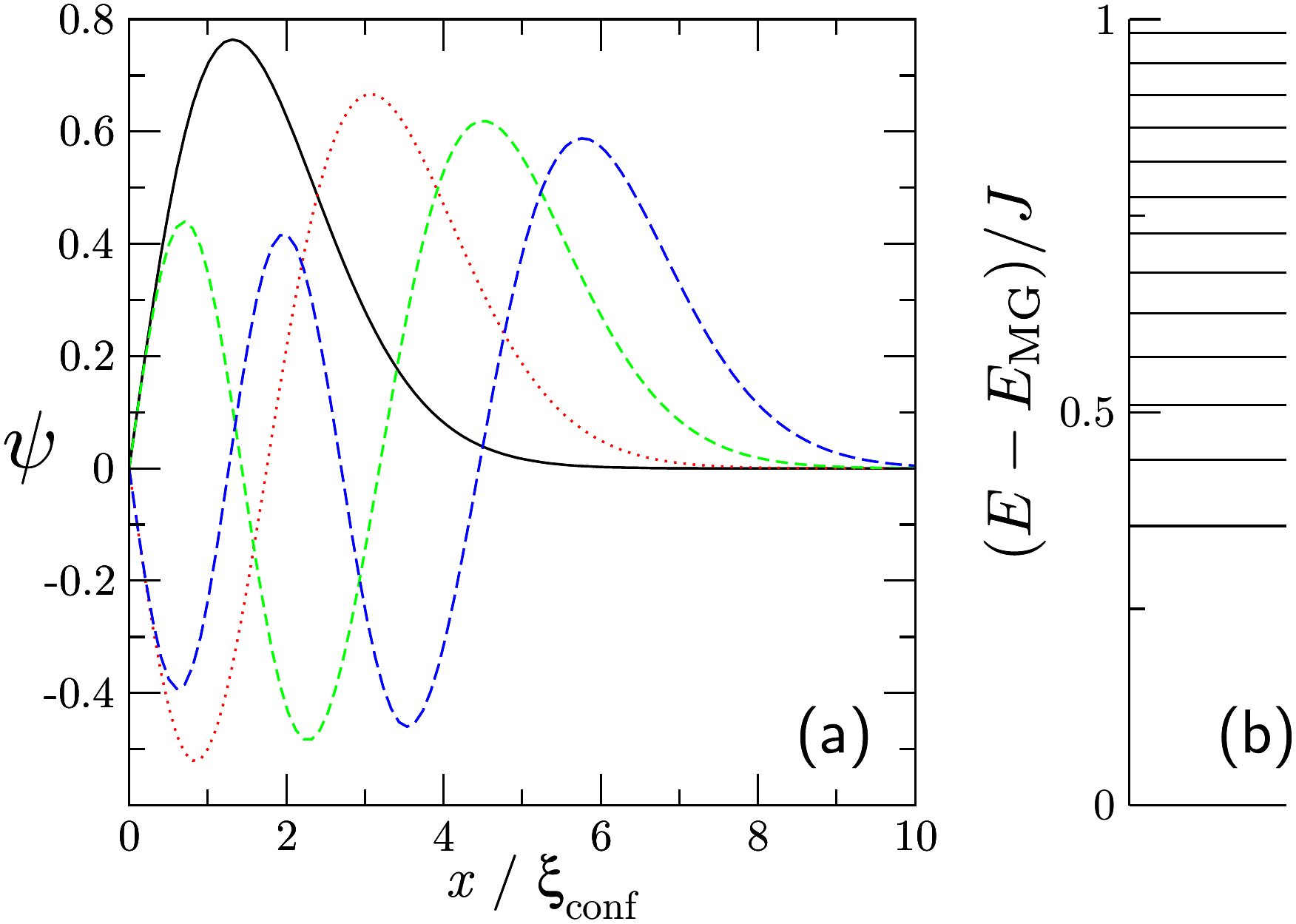}
\caption{(a) Airy functions corresponding to the first excited states confined close to $x=0$. (b) Corresponding energy levels for $\delta=0.01J$.}
\label{fig:Airy}
\end{figure}
The boundary conditions $\psi(x=-1)=0$ and
$\psi(x)\underset{x\rightarrow+\infty}{\longrightarrow}0$ allow the
following energies
\begin{equation}
 E_n=E_\text{MG}+\frac J4-\frac{3\delta}{4}(1+a_n\xi_\text{conf})\;,
\end{equation}
in which the $a_n<0$ are zeros of the $\Ai$ function. The associated
wave-functions read
\begin{equation}
\psi_n(x)\propto\Ai\left(\frac{x+1}{\xi_\text{conf}}+a_n\right)\;.
\end{equation}
These states and their energy are represented on Fig.~\ref{fig:Airy}.
Exact magnetization profiles obtained from DMRG have been compared to 
such analysis in Refs.~\cite{Uhrig1999} and \cite{Doretto2009}.

\subsection{Two spinons excitations}
\label{sec:etats_lies_delta}

In the case of an even chain with infinite number of sites and
containing two spinons, the explicit dimerization term yields a linear
(or string) attractive potential between the two domain walls. Writing
Schr\"odinger equation in the center of mass frame and taking the
continuum limit leads to
\begin{equation}
-4J\cos(K)\partial_x^2\psi(K,x)+\frac{3\delta}4\left[ x-\varepsilon(K)\right]\psi(K,x)=0\;,
\end{equation}
where
\begin{equation}
\varepsilon(K)=\frac{4}{3\delta}\left(E-E_\text{MG}\right)-\frac{2J}{3\delta}\left(5-4\cos K\right)\;.
\end{equation}
Using the change of variables
\begin{equation}
y=\frac{x-\varepsilon(K)}{\xi_\text{conf}(K)}\;,\quad\text{with}\quad
\xi_\text{conf}(K)=\left(\frac{16J}{3\delta}\cos K\right)^{1/3}\;,
\end{equation}
one recovers \eqref{eq:equadiff-Airy}.  When $K\rightarrow\pi/2$,
$\xi_\text{conf}\rightarrow0$ and taking the continuum limit is no
longer justified. However, for $K=\pi/2$, Schr\"odinger equation
simply reads
\begin{equation}
\left(E-E_{\text{MG}}\right)\psi_i=\left(\frac52J+\frac32\delta i\right)\psi_i\;.
\end{equation}
So the eigenstates are simply states with spinons at constant relative
distances $x=2i+1$. The corresponding energies are given by
\begin{equation}
\omega_i(\pi/2)=\frac52J+\frac32\delta i\;,\quad(i>0)\;.
\end{equation}
In the triplet sector, the bound-state with $i=0$ is still there,
which energy $\omega_0(\pi/2)\simeq2J$ is given by the boundary
condition \eqref{eq:limite_tp}.
\begin{figure}[t]
\centering
\includegraphics[width=0.8\linewidth,clip]{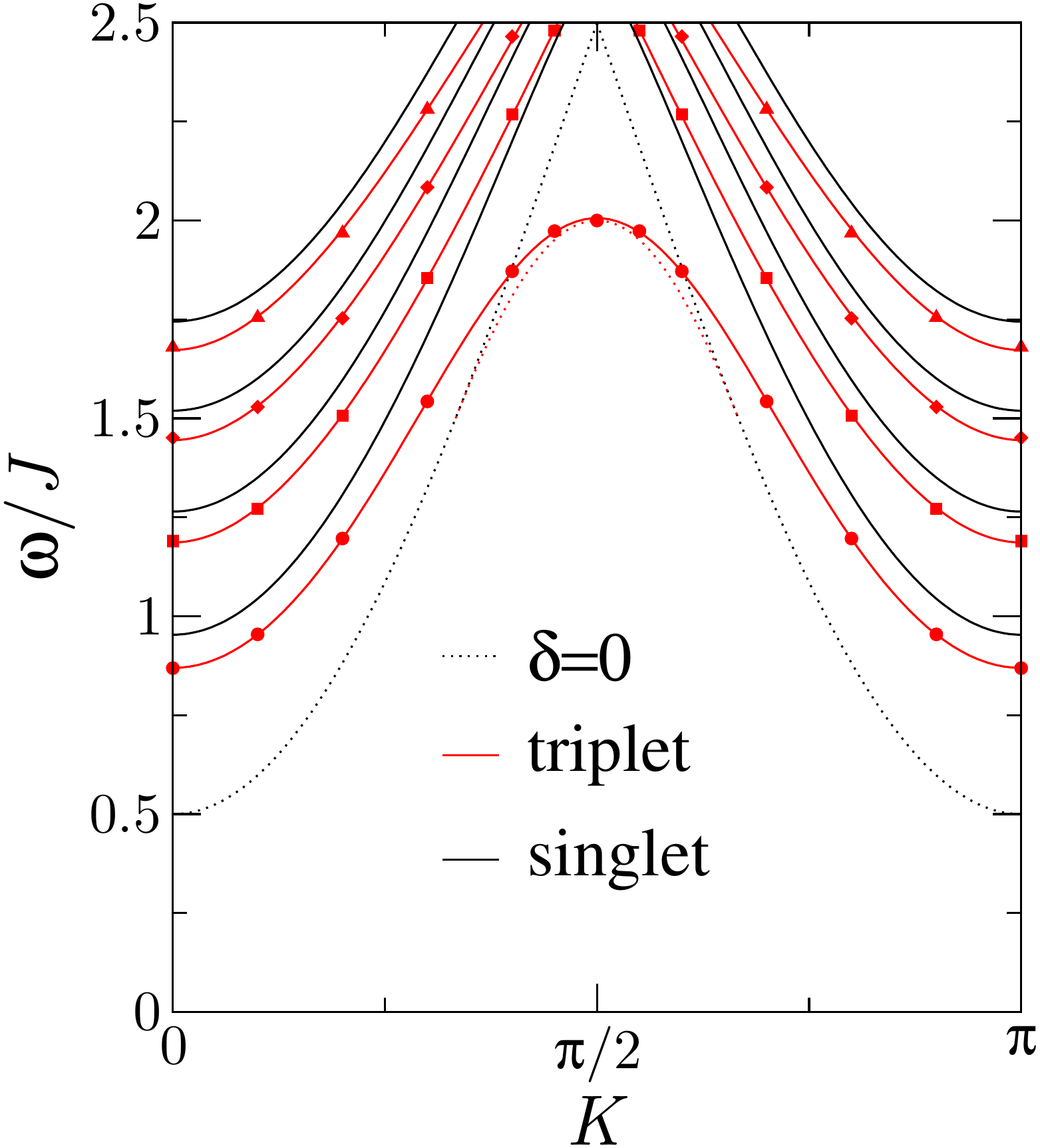}
\caption{Low-energy spectrum of the MG chain with explicit
  dimerization $\delta=0.05J$, as computed from RVB variational
  calculations in the continuum limit. In the triplet sector, symbols
  stand for numerical solutions of \eqref{eq:limite_tp_Airy} and thick
  lines are guide to the eyes.}
\label{fig:dispersion_delta}
\end{figure}
For $K<\pi/2$, the eigenstates are shifted Airy functions
\begin{equation}
\psi(K,x)\propto\Ai\left(\frac{x-\varepsilon(K)}{\xi_\text{conf}(K)}\right)\;,
\end{equation}
meaning that these are bound-states with characteristic length
$\xi_\text{conf}(K)$. Energies are determined from boundary
conditions \eqref{eq:limite_sg} in the singlet sector, and
\eqref{eq:limite_tp} in the triplet sector.  The singlet sector
directly gives the dispersion relation as a function of the zeros
$a_n$ of the Airy function $\Ai$:
\begin{equation}
\omega_n(K)=J\left(\frac52-2\cos K\right)+\frac{3\delta}4\left[1-a_n\xi_\text{conf}(K)\right]\;.
\end{equation}
In the triplet sector, the implicit equation on the energy $E$ can be
rewritten as
\begin{multline}
\label{eq:limite_tp_Airy}
-2\cos(K)\Ai\left(\frac{3-\varepsilon(K)}{\xi_\text{conf}(K)}\right)=\\
\left(\frac{3}{4}\frac{\delta}{J}\varepsilon(K)-\left(\frac12+2\cos K+\cos2K\right)\right)\Ai\left(\frac{1-\varepsilon(K)}{\xi_\text{conf}(K)}\right),
\end{multline}
which we solve numerically to obtain the spectra $E_n(K)$ and the
corresponding dispersion relations $\omega_n(K)$.
On Fig.~\ref{fig:dispersion_delta}, the first triplet and singlet
dispersion relations are displayed.  The lowest part of the continuum
\eqref{eq:dispersion_MG} now divides into single triplet and singlet
branches. These formation of bound-states can be also studied from a
weakly coupled dimer picture ($J_1-\delta\ll J_1$) followed by series
expansion of the coupling~\cite{Zheng2001}.  The behavior is
consistent with numerical studies of the elementary excitations of
this model~\cite{Sorensen1998,Bouzerar1998,Shevchenko1999,Singh1999}.

\section{Conclusion}

We obtained effective Schr\"odinger equations for the motion of a
single or two spinons in the frustrated chain by using the inverse of
the overlap matrix of the short-range RVB basis. In some peculiar
cases, the equations can be solved analytically, recovering known
results and providing new ones on the excitation spectra. ED results
are well accounted for, even for four-spinons continuum, assuming an
independent spinon picture. This approach is somewhat systematic as it
does not depend on the Hamiltonian, yet it works better when the
low-energy physics is well captured by dimer states. For instance, it
gave quantitative predictions in the random MG
model~\cite{Lavarelo2013}.

Frustrated chains are still an active field of experimental
research~\cite{Gibson2004,Schapers2013,Stone2014} and this approach
could shed light on related models. Furthermore, it would be
interesting to seek generalizations of the MG physics to higher spins,
as proposed in \cite{Rachel2009,Michaud2012,Matera2014} and to see
whether such approach could be also generalized. In particular, the
case of spin $3/2$ which is now relevant
experimentally~\cite{Damay2010,Damay2011} displays a qualitatively
similar phase diagram~\cite{Roth1998} as spin-$1/2$, yet with the
possibility to stabilize chiral phases in the presence of
anisotropy~\cite{Lecheminant2001,Hikihara2001}. Such model would
constitute a first step towards a generalization.

\subsection*{Acknowledgements}

We thank Matthieu Mambrini for insightful discussions. We acknowledge
support from the French ANR program ANR-2011-BS04-012-01 QuDec.

\end{document}